\documentclass{llncs}
\usepackage[utf8]{inputenc}
\usepackage{amsmath}
\usepackage{xcolor}
\usepackage{a4wide}
\usepackage{tablefootnote}
\usepackage{epsfig}
\usepackage{graphicx}
\usepackage{enumitem}
\usepackage{float}
\usepackage{url}
\usepackage{cite}

\usepackage{geometry}
\title{Neural Network Training on Encrypted Data with TFHE}
\twocolumn 

\author{Luis Montero, Jordan Frery, \\
        Celia Kherfallah, Roman Bredehoft, Andrei Stoian}
 \institute{Concrete-ML team, Zama} 

\begin{document}

\maketitle

\begin{abstract}
We present an approach to outsourcing of training neural networks while preserving data confidentiality from malicious parties.  We use fully homomorphic encryption  to build a unified training approach that works on encrypted data and learns quantized neural network models. The data can be horizontally or vertically split between multiple parties, enabling collaboration on confidential data. We train logistic regression and multi-layer perceptrons on several datasets. 
\end{abstract}

\section{Introduction}

Neural networks (NN) are powerful and versatile machine learning models that are the inspiration for state-of-the-art large language-models and deep convolutional neural networks. Training data is expensive to acquire and annotate, motivating data owners to protect it against attackers. Data confidentiality is maintained when training NNs locally, but requires data science know-how. Collaborative training NNs on data coming from multiple parties exposes data to a risk of leakage. This danger may be mitigated by using federated learning (FL), which is a multi-party protocol with a central server that performs gradient aggregation and model updates. However, the FL protocol requires active client participation and can not be applied to data that is stored encrypted without decryption. Furthermore, \emph{vertical} FL - a setting where different parties own different data features - can be complex to implement, and requires an additional, separate step to align the data between parties. The alignment also needs to be protected against leakage. Working with encrypted data end-to-end alleviates the operational complexity by using FHE as a uniform data processing framework.

We present a unified approach that can work without leakage with one or multiple parties, with both horizontal and vertical splitting. We use TFHE\cite{CGGI16} and we show how to accelerate encrypted quantized NN training using a recently introduced \emph{rounding} operator.

\section{Prior work}

\subsection{Encrypted training}
Logistic Regression training on encrypted data was described in several works \cite{Kim2018, han2019logistic, bergamaschi2019homomorphic, Bonte2018}. \cite{Bonte2018} uses the somewhat homomorphic FV scheme \cite{FV12} to implement a 2-nd order method with a ``fixed hessian" approximation. \cite{Kim2018} use CKKS and implement Nesterov accelerated gradient for training. \cite{han2019logistic} scales up this method to larger datasets by packing mini-batches in single ciphertexts and by applying boostrapping on weights every few iterations. All of these works use either polynomial or Taylor-series approximations of the sigmoid. 

For small multi-layer perceptrons (MLPs), \cite{lou2020glyph, nandakumar2019towards} present approaches that use SGD to train one and two-layer MLPs, showing results on MNIST or in transfer-learning. Both approaches use the BGV \cite{BGV14} scheme while \cite{lou2020glyph} adds  scheme-switching BGV-TFHE to accelerate Relu and SoftMax computation.  Both works use 80-bit security parameters, integer representations and they quantize weights, gradients, activations and the error function to 8-bits. 

\subsection{Integer arithmetic quantized training}
Training algorithms based on integer arithmetic are adapted to the encrypted training setting since TFHE is an integer scheme. \cite{wu2018training} introduces WAGE, an integer-arithmetic training algorithm that uses unbounded quantization of weights, activations gradients and errors. To account for changes in the variance of weights during training, pre-determined scaling factors specific to individual layers are used. For error propagation during training, a strong quantization that preserves gradient direction is applied. WAGE produces 2-bit weights and uses 8-bit activations, errors and gradients during training. Other works, such as \cite{NITI, yang2019swalp} improve upon WAGE by quantizing weights to 8-bits but use a dynamic exponent to quantize activations and weights. Dynamic exponents require the computation of the range of values in a tensor during its quantization, which is expensive on encrypted data. \cite{yang2019swalp} is used for FHE training by \cite{lou2020glyph}.

\section{Method}
We implement a neural network training algorithm where data, gradients and weights are encrypted. We quantize the weights, gradients, activations and the error computation and represent them with low bit-width integers. TFHE supports arithmetic computation on integers and programmable boostrapping ($PBS(f,x)$) which applies arbitrary functions $f$ on encrypted values $x$ while reducing their noise. The cost of PBS is strongly related to the bit-width of the encrypted integers it processes. We describe in this section how quantization keeps intermediary values low, therefore decreasing FHE training latency.

Our approach can perform quantized training on most types of  NNs, including MLPs and Logistic Regression. While it should be possible to train convolutional NNs with our approach, they are beyond the scope of this work. Through the PBS mechanism, all popular activation functions, such as Relu, Sigmoid, and Gelu are supported.

\subsection{Training computation graph}
\label{sec:graph}
\begin{figure*}
    \centering
    \includegraphics[width=0.9\textwidth]{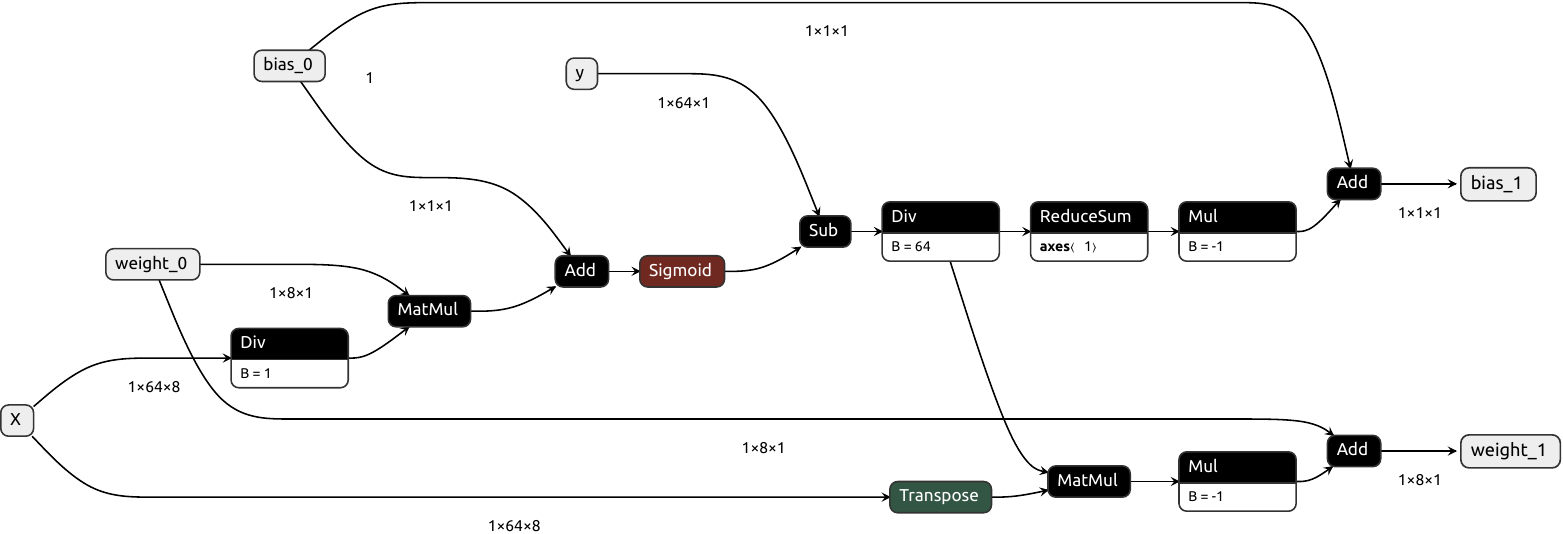}
    \caption{Training computation graph for Logistic Regression showing a single batch. \texttt{X} contains the training data $X$, \texttt{Y} the labels $y$. Initial trained parameters are stored in \texttt{bias\_0} (bias $b$), \textbf{weight\_0} (weights $w$). \texttt{MatMul}, \texttt{Add} and \texttt{ReduceSum} operations have calibrated input quantizers while \texttt{Mul}, \textbf{Div}, \texttt{Sigmoid} are computed with PBS.}
    \label{fig:graph}
\end{figure*}

We use PyTorch to express the code that performs training and ONNX to represent the computation graph that it produces. This approach is generic in terms of the model to be trained and allows us to use the same code when changing the model type or the gradient descent optimizer. In Figure \ref{fig:graph} we show the graph that performs the update of the weight and bias tensors for a single batch of training data for Logistic Regression. 

\subsection{Quantization}
Our work performs an off-line quantization calibration step using plaintexts. This step takes as input the floating point computation graph and a plaintext calibration data-set and produces an integer computation graph containing arithmetic operations and look-up table operations. First we compute statistics for each tensor that is processed in the graph. Based on these statistics the quantization parameters for each tensor are computed. Finally, the computation graph is modified by adding quantization and de-quantization functions before and, respectively, after every operation that works with encrypted integers in the ONNX graph. Quantization and activation functions are defined using floating point computations. During the calibration stage, all chained floating point operations, such as quantization and activations are converted to look-up table computations on integers. This fusion mechanism, described in \cite{stoian2023deep} does not incur any accuracy loss. 

The calibration stage makes some assumptions about the range of the values the training data ($X$) and trained parameters ($w,b$) can have. Calibration data is sampled as follows: $X, w, b \sim Uniform(-1, 1)$, $y \sim Bin(0.5)$. During the FHE training it is expected that the user provides training data with the same min/max values as the calibration data. After compilation, described in the next section, input data of the compiled quantized training graph (weights, biases, training data) must be  quantized before being encrypted.

\subsection{Compilation to TFHE}
\label{sec:compilation}
Next, we apply a compilation step to the computation graph produced by the previous step. The compilation produces machine code that accepts encrypted data and a PBS evaluation key and returns the encrypted weights after training on one data batch. The graph that is compiled contains:
\begin{enumerate}
    \item matrix multiplication (eq. \ref{eq:matmul}) and addition between encrypted and encrypted tensors, performing both PBS operations and arithmetic operations on ciphertexts

\small
\begin{equation}
\begin{gathered}
C_{ij} = \sum_k PBS(f^{sq}, A_{ik} + B_{kj}) - PBS(f^{sq}, A_{ik} - B_{kj}) 
\\ \quad
\forall i,j, f^{sq}(x) = \frac{x^2}{4}
\label{eq:matmul}
\end{gathered}
\end{equation}
\normalsize

    \item PBS operations that perform the table lookups corresponding to quantization and activation functions
\end{enumerate}
Our approach uses compilation, as described in \cite{Bergerat2023}. The computation graph is split into sub-graph partitions that contain one \emph{multi-sum} accumulation operation and a subsequent PBS operation. A \emph{multi-sum} is a series of additive arithmetic operations applied to ciphertexts. It can be, in the case of matrix multiplication, an accumulation of ciphertexts for each cell of the resulting matrix.

Crypto-system parameters are generated for each such partition and they are constrained to provide 128-bit security and to allow sufficient message space, in each graph partition, so that noise does not corrupt the accumulator of the \emph{multi-sum}. As several parameter sets are used in a single circuit, boostrapping keys must be generated for each parameter set. 

Applying this method, the accumulations performed by dense layers are fast levelled operations with correctness guarantees, at the expense of increasing the bit-widths of the subsequent PBSs. 

\subsection{TFHE Rounding operator}

The look-up tables computed with PBS in the compiled circuit contain a quantization function that reduces the bit-width of the accumulators by applying a scaling factor. Following \cite{jacob2018quantization} eq. 6, we express this scaling factor $M$ as:

\begin{equation}
 M = 2^{-n_{r}}M_0
\end{equation}

\begin{figure}
    \centering
    \includegraphics{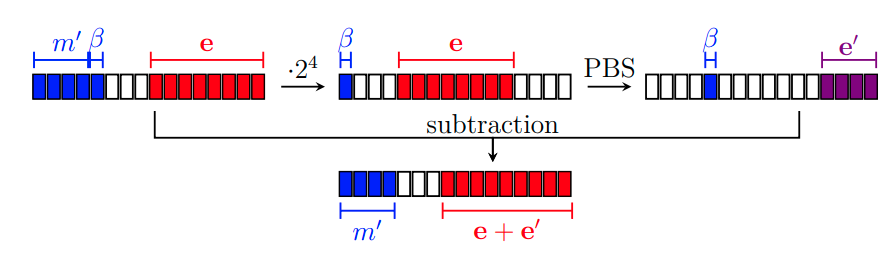}
    \caption{Rounded PBS removing the 1st LSB of a 5-bit value: first the least significant bit is shifted to become the most significant bit. Next, a 1-bit PBS moves the bit back to the 5-th position. A final subtraction removes the bit from the original value.}
    \label{fig:round_PBS}
\end{figure}

The $M_0$ value must remain large enough to prevent the multiplication by $M_0$ from degrading the mapping of inputs to their quantized versions. The quantizer inputs represent neuron outputs or gradients and are the accumulators of preceeding \texttt{MatMul} operations. We set $n_r$ experimentally. Once $n_r$ is set, computing the multiplication by $2^{-n_{r}}$ with TFHE is achieved using a bit-removal \emph{rounding} operator. Figure \ref{fig:round_PBS} shows the bit removal process. The PBS mentioned in the figure only bootstraps a single bit. 

\begin{table*}[t]
    \centering
    \begin{tabular}{|c|c|c|c|c|c|}
         \hline 
         Dataset & Model/\#params & Quantization $n$-bits & Best FHE Acc. & Best fp32 Acc. & Batch/Epoch latency \\
         \hline 
         breast-cancer & Logistic/30 & 4b & 98.25\% & 99.12\% & 11.8s / 0.23h\\
         breast-cancer & MLP (1 hidden)/930 & 4b & 98.25\% & 99.12\% & 149s / 2.94h\\
         mortality & Logistic/10 & 4b & 90.09\% & 90.47\% & 15.8s / 25.5h\\
         mortality & MLP (1 hidden)/165 & 4b & 87.25\% & 90.44\% & 45s / 72.78h\\
         \hline
    \end{tabular}
    \caption{Results for FHE Training on encrypted data. $\#params$ is the number of trainable parameters. Epoch latency is the time taken to train when each example in the data-set is used once. Latency is reported for a 16-thread execution on a 8-core processor.}
    \label{tab:results}
\end{table*}

\subsection{Mini-batch training and weight noise}

The computation graph in Figure~\ref{fig:graph} is exported for a single weight update based on one mini-batch and is compiled to a circuit as described in Section~\ref{sec:compilation}. 
After each mini-batch the weights are decrypted and re-encrypted to provide fresh encryptions for the next iteration of the mini-batch training circuit. 

This approach supposes that interaction occurs between the data owner and the server. It allows the data owner to evaluate model performance at each step of the training, allowing early-stopping. 
An alternative solution is to apply a bootstrapping to reduce the noise level in the encrypted weights after each mini-batch, making the whole training non-interactive. 

\section{Experimental Evaluation}

\subsection{Models and Datasets}
We test our method on two datasets: (1) \emph{mortality} from \cite{InfantMortalityDataset} containing 46582 examples with 10 features and two classes, (2) \emph{breast-cancer} from \cite{Street1993NuclearFE}\footnote{OpenML ID: 15} with 569 examples and 30 features.  We implement and evaluate accuracy and training latency per batch for $n=4$-bit logistic regression model and one-hidden layer MLP. We use a batch size of 8 for both datasets and a learning rate fixed at 1. 

\subsection{Results}

\begin{figure}
    \centering
    \includegraphics[width=0.5\textwidth]{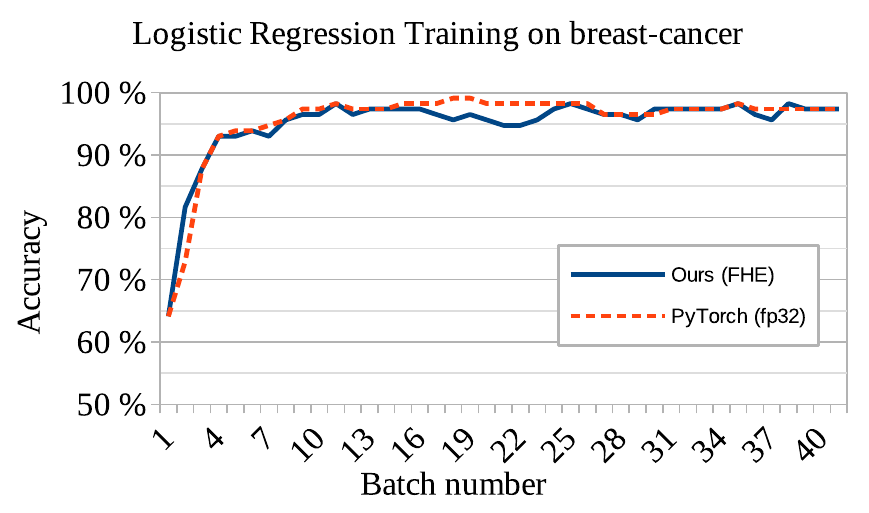}
    \includegraphics[width=0.5\textwidth]{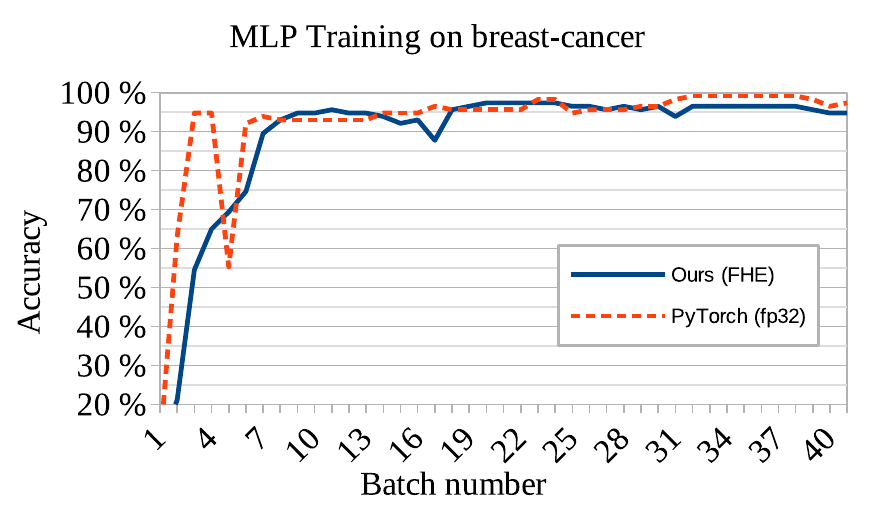}
    \includegraphics[width=0.5\textwidth]{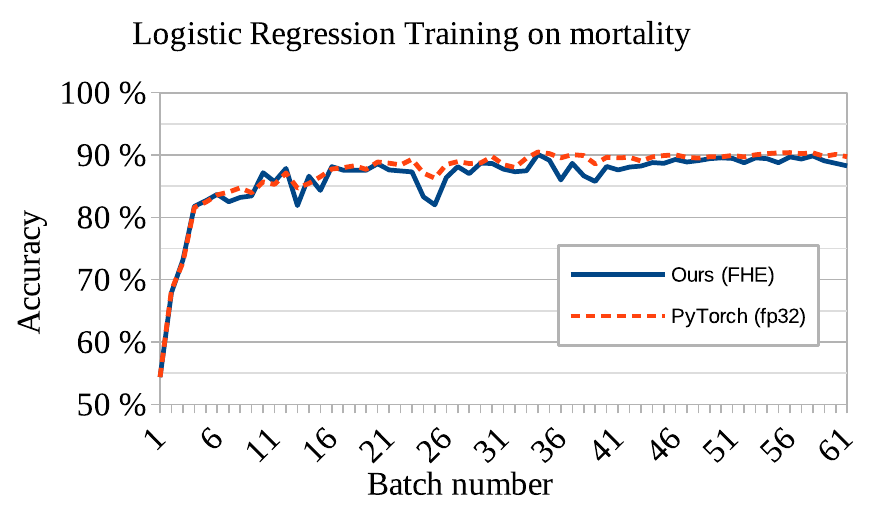}
    \includegraphics[width=0.5\textwidth]{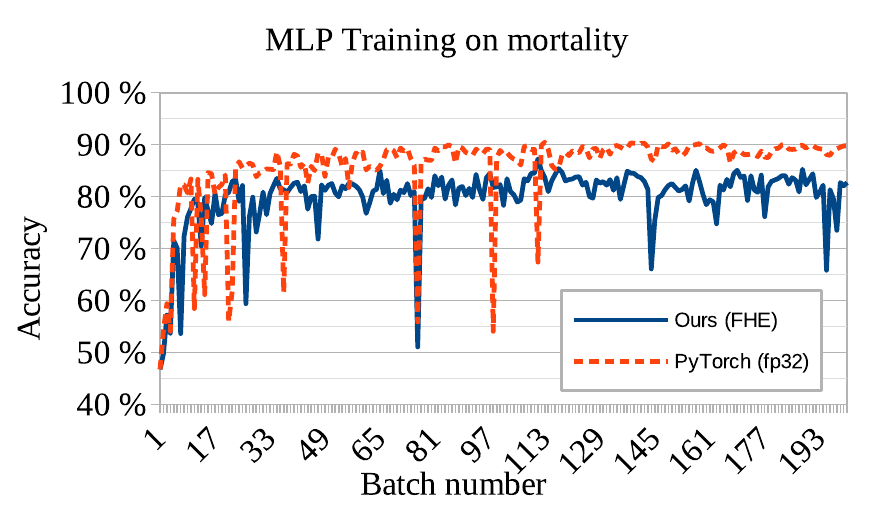}
    \caption{Accuracy during training for Logistic Regression and MLP on two datasets}
    \label{fig:results}
\end{figure}

Figure~\ref{fig:results} shows that training both of the encrypted models  converges robustly to the fp32 result on plaintexts. Accuracy is matched in all cases except for the MLP model on the more challenging \emph{mortality} dataset which shows a 2\% accuracy difference. Table~\ref{tab:results} summarizes the results and gives latencies for FHE training both models on both datasets. 

We extrapolate training latency results on MLPs to compare to \cite{lou2020glyph} and \cite{nandakumar2019towards}. To compare fairly, we define the individual weight gradient computations / second / thread \\ (WGC/s/T) unit. \cite{lou2020glyph} train a 2720-parameter MLP with a 60-example batch in 2.4 minutes using 48 threads which equates to 24 WGC/s/T. Based on our result on \emph{breast-cancer} in Table~\ref{tab:results}, we attain 3 WGC/s/T while  \cite{nandakumar2019towards} achieves only 0.4 WGC/s/T.

For Logistic Regression,  \cite{Kim2018, han2019logistic} use different datasets and fit the entire dataset in a single batch, making results not directly comparable. \cite{Kim2018} reach convergence after several epochs, while our approach, using mini-batches, converges faster. Overall their latency to convergence on data-sets of similar size to \emph{breast-cancer} is on the order of 3 minutes. Our mini-batch approach converges with 20 batches on \emph{breast-cancer} in a similar amount of time. 

\section{Conclusion}
We presented a novel encrypted neural network training method based on TFHE. It was demonstrated on logistic regression and on an MLP with one hidden layer. With 4-bit weights, gradients, activations and errors it trains models to accuracies on par with training on cleartext data for both logistic regression and MLPs. The latency of the training was optimized with a TFHE rounding operator that fits in naturally with neural network quantization.

\clearpage
\bibliographystyle{alpha}
\bibliography{main}
\end{document}